\documentclass[aps,prd,10pt,twocolumn,nofootinbib]{revtex4}

\usepackage{epsfig}
\usepackage{bm}
\usepackage{latexsym}
\usepackage{natbib}
\usepackage{url}
\usepackage{dcolumn}
\usepackage{color}
\usepackage{amsfonts,amssymb,amsmath}
\usepackage{graphicx,epsfig}
\usepackage{psfrag}
\usepackage{subfigure}
\usepackage{hyperref}
\hypersetup{colorlinks=true}
\usepackage{mathtools}
\usepackage{enumitem}

\def\gsim{\:\raisebox{-1.1ex}{$\stackrel{\textstyle>}{\sim}$}\:}
\newcommand{\ba}{\begin{array}}
\newcommand{\ea}{\end{array}}
\newcommand{\be}{\begin{equation}}
\newcommand{\ee}{\end{equation}}
\newcommand{\bea}{\begin{eqnarray}}
\newcommand{\eea}{\end{eqnarray}}

\def\beq{\begin{equation}}
\def\eeq{\end{equation}}
\def\bea{\begin{eqnarray}}
\def\eea{\end{eqnarray}}

\begin{document}

\title{Trans-Planckian Censorship Conjecture and Non-thermal post-inflationary history}

\author{
Mansi Dhuria$^{a,}$\footnote{mansidhuria@iitram.ac.in} and 
Gaurav Goswami$^{b,}$\footnote{gaurav.goswami@ahduni.edu.in}
}

\affiliation{
$^a$ Institute of Infrastructure, Technology, Research and Management, Ahmedabad 380026, India\\
$^b$ School of Engineering and Applied Science, Ahmedabad University, Ahmedabad 380009, India
}

\begin{abstract}
The recently proposed Trans-Planckian Censorship Conjecture (TCC) can be used to constrain the energy scale of inflation. The conclusions however depend on the assumptions about post-inflationary history of the Universe. E.g. in the standard case of a thermal post-inflationary history in which the Universe stays radiation dominated at all times from the end of inflation to the epoch of radiation matter equality, TCC has been used to argue that the Hubble parameter during inflation, $H_{\inf}$, is below ${\cal O}(0.1) ~{\rm GeV}$. 
Cosmological scenarios with a non-thermal post-inflationary history are
well-motivated alternatives to the standard picture and it is interesting to find out the possible constraints which TCC imposes on such scenarios. 
In this work, we find out the amount of enhancement of the TCC compatible bound on $H_{\inf}$ if post-inflationary history before nucleosynthesis was non-thermal. We then argue that if TCC is correct, for a large class of scenarios, it is not possible for the Universe to have undergone a phase of moduli domination.
\end{abstract}

\maketitle

\section{Introduction}

The most studied picture of the early Universe is that after the era of cosmic inflation, the Universe reheats, becomes radiation dominated, and stays radiation dominated all the way till Big Bang Nucleosynthesis (BBN). However, there are many reasons to entertain other possibilities for the history of post-inflationary Universe \cite{Kane:2015jia}. In particular, the dynamics and fate of moduli fields (such as those arising from string compactifications) in the early Universe have been studied for a very long time and in such scenarios, after the end of inflation and before the beginning of BBN, there can be an intermediate stage of deviation from radiation domination. 

The details of the dynamics of moduli fields depend crucially upon the details of cosmic inflation e.g. how the moduli masses compare with the energy scale of inflation etc. The observational upper limits on the tensor to scalar ratio, $r$, determine the upper limits on Hubble parameter during inflation $H_{\rm inf}$ or the scalar potential $V$ (see \cite{Akrami:2018odb} and first row of table \ref{table:summary_inf}). 
The lower limit on the energy scale of inflation arises from the requirement that the reheating temperature must be well-above that at the onset of BBN. Thus, there is a huge uncertainty in the energy scale of inflation and this translates into an uncertainty in our understanding of non-thermal cosmological scenarios.

If Trans-Planckian Censorship Conjecture (TCC) \cite{Bedroya:2019snp} is true, then, it can be used to put constraints on the energy scale of inflation \cite{Bedroya:2019tba}.
\footnote{See e.g. \cite{Cai:2019hge,Tenkanen:2019wsd,Das:2019hto,Mizuno:2019bxy,Brahma:2019unn} for other recent papers related to TCC.}
It turns out that the inferred constraint depends on the post-inflationary history of the Universe in a rather dramatic way (see e.g. the second and third row of table \ref{table:summary_inf}). 
E.g. in \cite{Bedroya:2019tba}, the authors assume that the post inflationary universe was radiation dominated (so that the equation of state is $p/\rho \approx 1/3$) and obtained the bound $r \le 10^{-30}$ which gets saturated for $H_{\rm inf} \approx 0.25$ GeV i.e. $V^{1/4} \approx 10^9$ GeV (see third row of table \ref{table:summary_inf}). 
In contrast, ref \cite{Mizuno:2019bxy} considered a less conservative evolution of the universe at the end of inflation and found that if the effective equation of state of the universe between end of slow roll and big bang nucleosynthesis is $w = p/\rho \approx -1/3$, then, one could have $r \le 10^{-8}$ and this saturates when $H_{\rm inf} \approx 10^{10}$ GeV i.e. $V^{1/4} \approx 10^{14}$ GeV (see second row of table \ref{table:summary_inf}).

 \begin{table}[h] \label{table:summary_inf}
\begin{tabular}{c c c c c c}
\hline
Sr. & $r$ & $H_{\inf} $ & $V^{1/4}$ & $w = \frac{p}{\rho}$ & slope  \\
No. &  $<$  &  (in GeV)  &  (in GeV)  &                             & $\frac{3}{2} (1+w)$\\
&&&&&\\
\hline
\hline
&&&&&\\
 1 & 0.07 & $10^{13}$ & $10^{16}$ & - & - \\
&&&&&  \\ 
 2 & $10^{-8}$ & $10^{10}$ & $10^{14}$ & $- \frac{1}{3}$ & 1 \\
&&&&&  \\ 
 3 & $10^{-30}$ & $10^{-1}$ & $10^{9}$ &  $+ \frac{1}{3}$ & 2 \\  
&&&&&  \\
 \hline 
\end{tabular}
\caption{In this table, $w=\frac{p}{\rho}$ is the equation of state parameter of the cosmic fluid which dominates the energy density of the universe after the end of slow-roll inflation and just before the beginning of BBN. In a plot of $\log L_{\rm phy}(t)$ vs $\log a(t)$, the physical size of Hubble radius $H^{-1}(t)$ is a straight line whose slope is determined by $w$ and this is given in the last column in the table.}
\end{table}

Since the value of $H_{\inf}$ for the two scenarios are so vastly different, the various details of phenomenology as well as those of the history of the Universe would be very different.
E.g. for $H_{\inf}$ as low as $10^{-1}$ GeV, the vacuum instability in Higgs potential at high scales will not be so problematic \cite{Buttazzo:2013uya,Espinosa:2015qea} (see also \cite{Goswami:2014hoa}).
This fact, and a look at the last two rows of table \ref{table:summary_inf} suggests that, for a Universe which undergoes a post-inflationary history during which, the equation of state parameter $w$ of the dominant component takes a value in between $-1/3$ and $1/3$, the constraints on $H_{\inf}$ would take some value in between $10^{-1}$ GeV and $10^{10}$ GeV. 

In particular, for a cosmology in which the post-inflationary history of the Universe is dominated by moduli, $w = 0$ and one expects that this will lead to enhancement of the upper limit on $H_{\inf}$ obtained using TCC (as compared to \cite{Bedroya:2019tba}). The actual amount of this enhancement can have though-provoking implications for moduli dominated cosmology. Recall that for moduli with gravitational interactions, requiring that the reheating temperature be above the temperature of BBN, the moduli mass $m_{\rm mod}$ must be greater than about $10$ TeV. One must also recall that a moduli dominated Universe behaves like matter dominated only when the moduli get displaced from the minimum of the potential and undergo oscillations.
But the modulus gets displaced from the minimum of its potential if $H_{\inf} > m_{\rm mod}$. Thus, if $H_{\rm inf} < m_{\rm mod}$ , then, the moduli will not be displaced from their potentials during inflation and hence there will be no moduli dominated phase.

Thus, it is important to know whether the enhancement in $H_{\inf}$ inferred from TCC, which will take place in a case in which post-inflationary history is matter dominated for some duration before BBN, will make it sufficiently bigger than about $10$ TeV or not.
In this paper, we answer this question by carefully finding the amount of enhancement in $H_{\inf}$ inferred from TCC for several choices of the duration of pre-BBN matter dominated phase. 

The paper is organised as follows:
in the next section i.e. \textsection \ref{sec:remind}, we remind the reader some very basic concepts about TCC and moduli dominated cosmology. Then, in \textsection \ref{sec:formalism}, we present the method we use to implement TCC to find the constraint on $H_{\rm inf}$ for every possible choice of duration of non-thermal history. 
In \textsection \ref{sec:results}, present the results we obtain and try to understand them in an approximate analytical way. Finally,
we conclude in \textsection \ref{sec:discussion} with a discussion about what one learns from the calculations of this paper.

\section{Trans-Planckian Censorship, inflation and cosmological moduli} \label{sec:remind}

In this section, we remind the reader some basic concepts relevant to the discussion in the rest of the article.

\subsection{Trans-Planckian Censorship Conjecture (TCC) and inflation}

Let $a_i$ be the value of scale factor at $t=t_i$, the beginning of inflation. 
There must be a length scale whose physical wavelength (at the instant $t_i$) is equal to Planck length.
The comoving wavelength of this mode would be
\begin{equation} \label{eq:pl_len_com}
 {\ell}_{co} = \frac{{\ell}_{pl}}{a_i} \; .
\end{equation}
Trans-Planckian Censorship Conjecture (TCC) \cite{Bedroya:2019snp,Bedroya:2019tba} demands that, the dynamics of the Universe must be such that none of the modes whose physical wavelength at $t_i$ is smaller than or equal to the physical wavelength of this mode, should ever become super-Hubble. 
The basic motivation for demanding this can be understood by recalling that, during inflation, as the wavelength of any Fourier mode of metric perturbation becomes much larger than Hubble radius, the mode function freezes and quantum fluctuations on that length scale become classical
\cite{Polarski:1995jg,Lyth:2006qz,Kiefer:2008ku}. Thus, TCC dictates that quantum fluctuations at sub-Planckian length scales should not become classical. 

At the moment when inflation ends, let the scale factor be $a_f$, then, the physical wavelength of the mode with comoving wavelength given by Eq (\ref{eq:pl_len_com}) will be
\begin{equation} 
 {\ell}_{phy} =  {\ell}_{co}  a_f  = \frac{{\ell}_{pl}}{a_i} a_f \; ,
\end{equation}
then, if the Hubble radius at the end of inflation is $H_f^{-1}$, then, TCC says that
\begin{equation} \label{eq:TCC}
 \frac{{\ell}_{pl}}{a_i} a_f < H_f^{-1} \; ,
\end{equation}
which is the form in which one can implement TCC.
 
Before proceeding, let us recall that, as of today, inflation gets observationally constrained by the values of only the following parameters: scalar spectral amplitude $A_s$, scalar spectral index $n_s$ and (upper limits on) tensor to scalar ratio $r$.
The energy scale of inflation could be characterised by either the Hubble parameter during inflation $H_{\rm inf}$, or the scalar potential $V$ or the field excursion (also, recall that the number of e-foldings of inflation required depends on the assumed energy scale of inflation). 
The energy scale of inflation has a large uncertainty: the upper limits come from the observational upper limits on $r$ and probably the lower limit comes from the requirement of not spoiling BBN.
In the simplest picture of inflation, $H_{\rm inf}$ and $V_{\rm inf}$ are related to observables $r, A_s$ in the following manner:
\begin{equation}
\left( \frac{H_{\rm inf}}{M_{pl}} \right)^2 = \frac{1}{3} \left( \frac{V^{1/4}_{\rm inf}}{M_{pl}} \right)^4 = \frac{\pi^2}{2} A_s r \; ,
\end{equation}
the excursion of inflaton during inflation i.e. $\Delta \phi$ is given by
\begin{equation}
\frac{\Delta \phi}{M_{pl}} = {\cal O}(1) \times \left( \frac{r}{0.01} \right)^{1/2} \; ,
\end{equation}
If TCC is true, Eq(\ref{eq:TCC}) can be used to constrain the energy scale of inflation \cite{Bedroya:2019tba}. 
But as we argued (and as we shall see in greater detail), the energy scale of inflation inferred from TCC depends on the post-inflationary history of the Universe. For a standard post-inflationary history of the Universe, the TCC implies a very low scale inflation (in this context, see \cite{Knox:1992iy}).

\subsection{Moduli and cosmology}

Moduli are scalar fields generically present in consistent 4-D solutions of string theory. They are massless at some leading order of description but their potential gets generated and stabilised by various effects such as fluxes, perturbative/non-perturbative corrections and/or SUSY breaking effects etc. present in specific string constructions~\cite{Kachru:2003aw,Balasubramanian:2005zx,Acharya:2008zi}. When these effects are taken into account, they acquire masses of the order of SUSY breaking scale or larger than that (except axions which remain light due to their shift symmetry). 

During inflation, a modulus whose mass $m_{mod} \ll H_{inf}$ will get displaced from the minimum of its potential ~\cite{Dine:1995kz}. 
When the Hubble scale becomes equal to the mass of the moduli, these fields start to oscillate around their post-inflationary minima with an initial amplitude given by the difference between the inflationary and the post-inflationary low energy minima of the moduli. 
This leads to an epoch in the history of the universe in which the energy density of the post-inflationary universe is dominated by coherent oscillations of the moduli fields. 
The shape of the scalar potential of the modulus field at its minimum, determines the effective equation of state parameter $w = p/\rho$ \cite{Turner:1983he} in this intermediate stage. 
The most robust constraint on this possibility comes from the fact that this modified history of the Universe should not spoil predictions of BBN \cite{Kawasaki:1999na}.

As $H$ decreases and becomes of order of the decay width of the moduli ($\Gamma_{\rm mod}$), they decay and the universe again enters the radiation dominated era. The decay of moduli into relativistic SM particles and/or other light relativistic degree of freedom will increase the entropy of the universe, thus reheating the universe again. The requirement that the decay of the moduli into SM particles shall not spoil the constraints
imposed on the abundances of light elements produced by BBN leads to the condition that the reheating temperature of universe after the decay of modulus is greater than ${\cal O}(\rm MeV)$. 
For moduli with gravitational couplings, 
\begin{equation}
 \Gamma_{\rm mod} \sim \frac{m_{mod}^3}{M_{pl}^2} \; ,
\end{equation}
and since 
\begin{equation}
 H \sim  \left(\frac{\pi^2 g_*(T)}{90}\right)^{1/2} \frac{T^2}{M_{pl}} \; ,
\end{equation}
in radiation dominated era (here $g_*(T) \approx 100$ is the relativistic degree of freedom contributing to the energy density). the condition $H = \Gamma_{\rm mod}$ and the fact that $T \ge T_{bbn} \approx {\rm MeV}$ tells us that 
 \begin{equation}
  m_{\rm mod} \ge  10~{\rm TeV}. 
 \end{equation}
Thus considerations based on BBN lead to a lower bound on moduli masses, known as ``cosmological moduli problem" bound. 
Given all of these details, the question arises, if we consider a non-thermal post inflationary history and then determine the constraint which TCC imposes on $H_{\inf}$, can $H_{\inf}$ ever be larger than the smallest possible values of masses of moduli? 

\section{Strategy of calculation} \label{sec:formalism}

The question we wish to ask is, if we now introduce an era of moduli domination, to what extent does the constraint on Hubble parameter during inflation change.
For the case for which $w \approx -1/3$, ref \cite{Mizuno:2019bxy} presented a clever argument to quickly determine $H_{\inf}$. 
Here we would like to take a very conservative approach: after the end of slow-roll inflation and before the beginning of Big Bang Nucleosynthesis, there is a some duration for which the Universe is matter dominated i.e. $w \approx 0$. Since we are dealing with a slightly more complicated history of the Universe, we need to implement TCC  numerically to find possible constraints on $H_{\inf}$. 
As we shall see, increasing the duration of matter domination will lead to an increase in the TCC compatible upper-limit on $H_{\inf}$.

\subsection{Basic formalism}

\begin{figure*}
  \includegraphics[width = 0.95\textwidth]{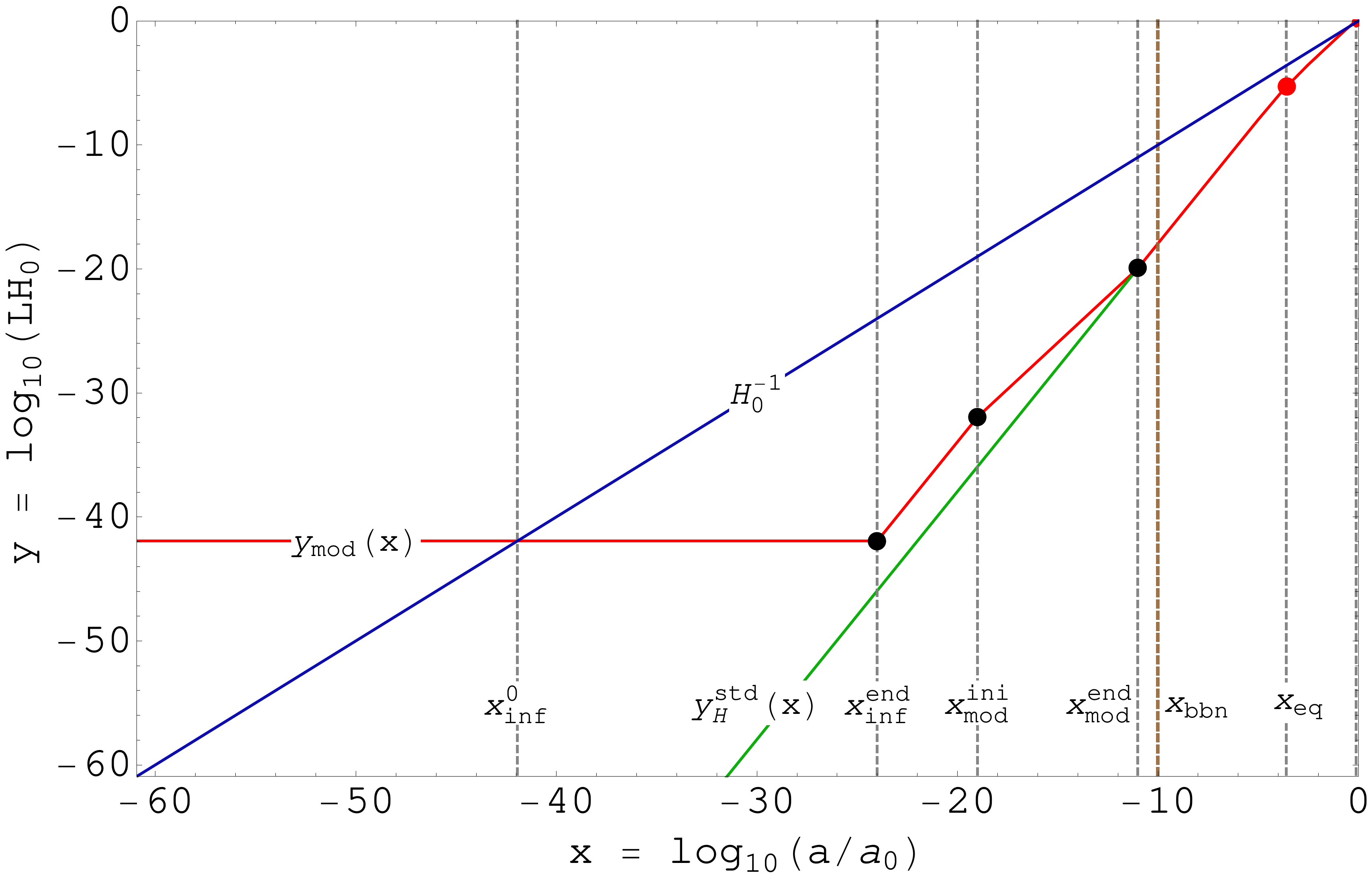}
  \caption
 {This figure provides a summary of all the quantities presented in the paper. As expected, $y_H^{std} (x)$ and $y_{mod}(x)$ are identical for $x \geq x_{mod}^{end}$, so the green line sits below the red line for $x>x_{mod}^{end}$ and hence can't be seen.}
  \label{evol-simple}
\end{figure*}

To have a sufficiently robust approach to constraining $H_{\inf}$ using TCC, we need to a way to numerically implement all the ideas. We now describe the details of our approach. As we shall see, we still need to make a number of concrete simplifying assumptions which we'll state later in the section. Let us define 
\begin{equation}
x = \log \left( \frac{a}{a_0} \right) \; ,
\end{equation}
here, $a$ is the scale factor at any epoch and $a_0$ is scale factor now. Note that all the logarithms, unless otherwise stated, are to base 10.
Similarly, define
\begin{equation}\label{eq-def-y}
y = \log \left( \frac{L}{H_0^{-1}} \right) \; ,
\end{equation}
here, $L$ is any length scale of interest and the argument of the log is the length scale $L$ in units of Hubble distance today.
We shall be interested in the range of length scales from Planck length ${\ell}_{pl}$ to $H_0^{-1}$ (the Hubble distance today).
At this stage, it is useful to have a way to visualise what we are doing, thus, it is advisable that the reader keeps looking at fig (\ref{evol-simple}) while reading this section.
Thus, the range of $y$ values of interest will be
\begin{equation}
y_{min} = \log \left( \frac{{\ell}_{pl}}{H_0^{-1}} \right) \; , {\rm and},~  y_{max} = 0 \; .
\end{equation}
Similarly, the range of $x$ values would be $x_{min} = y_{min}$ and $x_{max} = 0$.
This implies that $x_{min} = y_{min} \approx -60$ (see the axes in fig (\ref{evol-simple})).

Since the Universe is expanding, given any length scale today, its physical length at some time in the past would be smaller. Consider the length scale which today has length equal to $H_0^{-1}$, its $y$ value at any time in the past (specified by $x$) would be
\begin{equation}
y_{H_0} (x) = H_0^{-1} x \; .
\end{equation}
This is represented as the blue line marked ``$H_0^{-1}$" in fig (\ref{evol-simple}).
Let $y_{H}^{std} (x)$ be the $y$ coordinate corresponding to Hubble parameter in the standard model of cosmology (i.e. without inflation or moduli domination etc). This quantity (which is just log of Hubble distance at any time in units of $H_0^{-1}$) evolves as
\begin{equation} \label{yHstd}
y_{H}^{std} (x) = \log \left[ \Omega_{\Lambda} + \Omega_{k} 10^{-2x} + \Omega_{m} 10^{-3x} +  \Omega_{r} 10^{-4x} \right]^{ -\frac{1}{2} } \; ,
\end{equation}
where, the $\Omega$s are various density parameters today and hence $ \Omega_{\Lambda} + \Omega_{k} + \Omega_{m} +  \Omega_{r} = 1$.
Note the green line in fig (\ref{evol-simple}) marked ``$y_{H}^{std} (x)$" (note that, in the figure, for $x>x_{\rm mod}^{\rm end}$, the green line can not be seen because it sits below the red line, so, is hidden from view).
Even in this standard big bang model, there are special epochs e.g. the epoch of matter-radiation equality
\begin{equation}
 x_{eq} = \log \frac{\Omega_r}{\Omega_m} \; ,
\end{equation}
the epoch of the beginning of late time acceleration of the Universe
\begin{equation}
 x_{de} = \frac{1}{3} \log \frac{\Omega_m}{\Omega_\Lambda}  \; .
\end{equation}
Before proceeding, we must mention that for Big Bang Nucleosynthesis, 
\begin{equation}
\frac{a}{a_0} = \frac{T_0}{T} \approx 10^{-10} \; ,
\end{equation}
so that this epoch corresponds to $x_{bbn} \approx -10$.
All these three special epochs are represented by vertical dashed lines marked by their names in fig (\ref{evol-simple}), in particular, the line corresponding to $x_{bbn}$ is of brown color.

\subsubsection{Including moduli dominated phase}

In order to analyse the behaviour of the Universe with a moduli domination phase, we shall consider a scenario in which the history of the Universe for $x > x_{eq}$ is the same as that in standard big bang cosmology but before that, the early history gets modified in the manner summarised in Table \ref{table:summary}. Thus, as compared to standard big bang cosmology, we have introduced the following in the early universe: 
for $x < x_{inf}^{end}$, the universe undergoes cosmic inflation, after which the Universe gets dominated by radiation,
at $x = x_{mod}^{ini}$, the Universe gets dominated by moduli, this ends at $x_{mod}^{end}$ after which the radiation domination, and hence the usual thermal history starts. Thus, we must have
\begin{equation} \label{ineq_all_xs}
 x_{inf}^{end} < x_{mod}^{ini} < x_{mod}^{end} < x_{bbn} < x_{eq} \; ,
\end{equation}
Obviously, we have to have $x_{mod}^{end} < x_{eq}$, but note that we must also have $x_{mod}^{end} < x_{bbn}$.
All of these additional epochs are represented by additional vertical dashed lines marked by their respective names in fig (\ref{evol-simple}).

 \begin{table}[h] 
\begin{tabular}{cll} 
\hline
Sr. & Range of $x$ & What happens  \\
No. &&\\
\hline
\hline
  & & \\
 1  & $x < x_{inf}^{end}$ & ~~inflation \\ 
  & & \\
 2  & $x_{inf}^{end} <  x < x_{mod}^{ini} $ & ~~radiation domination \\ 
   & & \\
 3  & $x_{mod}^{ini} <  x < x_{mod}^{end} $ & ~~moduli domination \\ 
   & & \\
 4  & $x_{mod}^{end} <  x < x_{eq} $ & ~~radiation domination \\ 
   & & \\
 \hline 
\end{tabular}
\caption{A summary of the history of the Universe we are interested in.}
\label{table:summary}
\end{table}

At the epoch of the end of moduli domination, with $x=x_{mod}^{end}$, and we must have
\begin{equation}
 \Omega_{\rm mod} = \Omega_r 10^{-x_{mod}^{end}} \; ,
\end{equation}
where $ \Omega_{\rm mod}$ is the density parameter of moduli, similarly, at the epoch of beginning of moduli domination,
\begin{equation}
\Omega_r^E =  \Omega_{\rm mod} 10^{x_{mod}^{ini}} \; ,
\end{equation}
here, $\Omega_r^E$ is the density parameter corresponding to the radiation in the early Universe before moduli domination began.
It is now easy to see that for this case which involves inflation as well as a phase of moduli domination, the equivalent of Eq (\ref{yHstd})
shall be 

\begin{equation}
\label{eq:main}
y_{\rm mod}(x) = 
    \begin{dcases}
        y_H^{std}(x) \; , & x \geq x_{mod}^{end} \\
        \log \left( \Omega_{\rm mod} 10^{-3x} \right)^{-\frac{1}{2}} \; , & x_{mod}^{end} \geq x \geq x_{mod}^{ini} \\
        \log \left( \Omega_{r}^E 10^{-4x} \right)^{-\frac{1}{2}} \; , & x_{mod}^{ini} \geq x \geq x_{inf}^{end} \\        
        \log \left( \Omega_{r}^E 10^{-4 x_{inf}^{end}} \right)^{-\frac{1}{2}} \; , & x_{inf}^{end} \geq x  \\        
    \end{dcases}
\end{equation}

This function $y_{\rm mod}(x)$ is a logarithmic measure of Hubble distance at any time (in units of Hubble distance now) and is represented by red line in fig (\ref{evol-simple}). The thick dots on the red line in fig (\ref{evol-simple}) represent the epochs at which the behaviour $y_{\rm mod}(x)$ transitions.
Before proceeding, let us note that we have kept $y_{\rm mod}(x)$ for $ x \leq x_{inf}^{end} $ to be a constant. We are thus working with the approximation that the Hubble parameter during inflation is treated as exactly constant. At a later state, we'd comment on how the conclusions we draw could change if we do not make this assumption.

\subsubsection{Implementing TCC}

In order to implement this, we would make two additional assumptions:
\begin{enumerate}
 \item At the time $t = t_i$, the mode corresponding to current Hubble radius is just entering the Hubble sphere, i.e. inflation is ``just enough"
 (note that this is epoch is shown by a vertical line marked ``$x_{\inf}^0$"),
 \item The Fourier mode whose comoving wavelength is given by Eq (\ref{eq:pl_len_com}) must saturate the TCC bound Eq (\ref{eq:TCC}). 
\end{enumerate}
This suggests the following way of implementing TCC: at the time $t_i$ (with $x = x_{\inf}^0$), consider the mode whose physical wavelength at this time is equal to ${\ell}_{pl}$, this mode, must be just inside the Hubble radius at the end of inflation. Note that such a mode could be represented as a line parallel to the blue line in fig (\ref{evol-simple}) and passing through the point $(x_{\inf}^0,y_{min})$. 
Keeping this picture in mind, one realises that once the above assumptions are made, it is easy to see how TCC can be implemented numerically to constrain $H_{\inf}$ for any choice of $x_{mod}^{end}$ and $x_{mod}^{ini}$. We have implemented the following algorithm: for every choice of values of $x_{mod}^{end}$ and $x_{mod}^{ini}$ satisfying Eq (\ref{ineq_all_xs}),
\begin{enumerate}[label=(\alph*)] 
 \item Choose a large enough $x_{inf}^{end}$ (which still satisfies Eq (\ref{ineq_all_xs})),
 \item Find $y_{\rm mod}(x)$ at $x = x_{inf}^{end}$ using Eq (\ref{eq:main}),
 \item Now find the value of $x$ at the epoch $t_i$ when $x = x_{\inf}^0$ i.e. when the mode corresponding to current Hubble radius exits the Hubble radius during inflation, this can be done by solving for $x$ in the equation $y_{H_0} (x) =  y_{\rm mod}(x_{inf}^{end})$, call the solution of $x$ for this Eq to be $x_{\inf}^0$, note that this is where we are making the assumption that Hubble parameter during inflation is a constant,
 \item Find the mode whose physical wavelength at this epoch $x_{\inf}^0$ is equal to Planck length: such a mode would be described the point $(x_{\inf}^0,y_{min})$, 
 \item Find the physical wavelength of this mode at any time, this will be described by the equation 
 \begin{equation}
  y = y_{min} + ( x - x_{\inf}^0) \; ,
 \end{equation} 
 \item Using this, find the physical wavelength of this mode at the end of inflation i.e. set $x$ to be $x_{inf}^{end}$ in the above Eq and solve for $y$, call the corresponding value of $y$ to be $y_f$
 \item if $y_f > y_{\rm mod}(x_{inf}^{end})$, then TCC is not satisfied, in this case, go back to step 1 above and try a smaller value of $x_{inf}^{end}$. On the other hand, if $y_f < y_{\rm mod}(x_{inf}^{end})$, then TCC is satisfied. Since we wish to saturate the TCC bound, this is not good enough, we must go back to step 1 to try a large value of $x_{inf}^{end}$ until we find $y_f > y_{\rm mod}(x_{inf}^{end})$ within some accuracy.
\end{enumerate}

Once we follow this procedure, for every choice of values of $x_{mod}^{end}$ and $x_{mod}^{ini}$, we can find $x_{inf}^{end}$. From this, we can get $y_{\rm mod}(x_{inf}^{end})$ and this can be used to find the Hubble parameter during inflation in units of Hubble parameter today using Eq (\ref{eq-def-y}).

\section{Results}
\label{sec:results}

\subsection{Numerical results}

\begin{figure}
  \includegraphics[width = 0.45\textwidth]{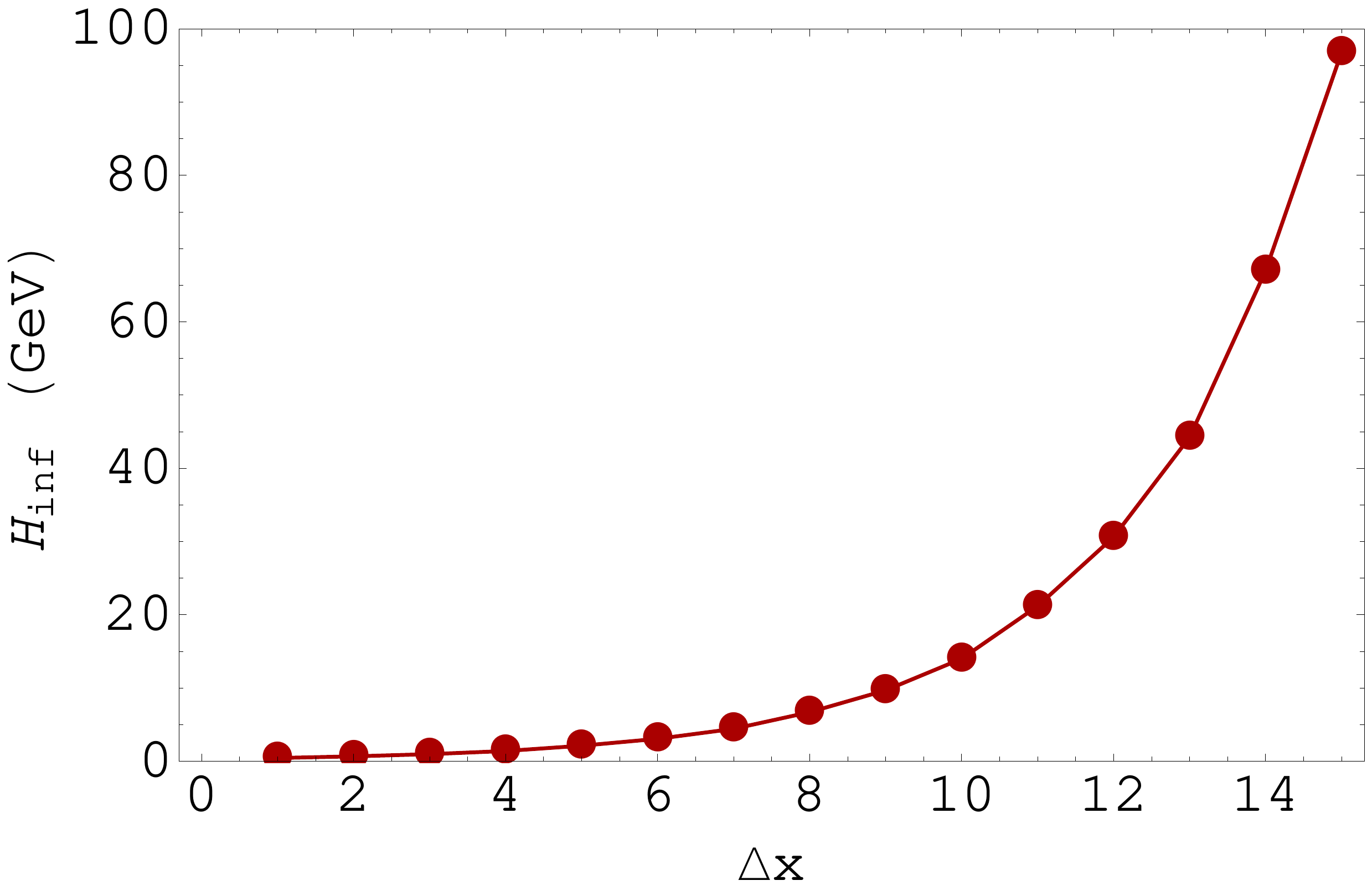}
  \caption
 {The Hubble parameter during inflation, $H_{\inf}$, as determined by implementing the procedure described in the text (and based on TCC) as a function of $ \Delta x$ which is a logarithmic measure of the duration of moduli domination in the early Universe. The presence of moduli dominated phase does increase the inferred $H_{\inf}$ as compared to a case in which post-inflationary history is purely thermal.
 }
  \label{Hvsdx}
\end{figure}

We set the values of cosmological parameters to their best fit values\cite{Akrami:2018odb} and then, for every choice of $x_{mod}^{end}$ and $x_{mod}^{ini}$, we find $y_{mod}(x)$ and use the procedure described by the last section to find TCC compatible $y_{mod}(x_{inf}^{end})$. 
We thus have two free parameters, and we note that, once we fix $x_{mod}^{end}$ and $x_{mod}^{ini}$, when we use TCC to fix $H_{\inf}$, we do not have a choice in deciding how much is the number of e-foldings of expansion of the Universe between the end of inflation and beginning of moduli domination. Let us now define
\begin{equation}
 \Delta x = x_{mod}^{end} - x_{mod}^{ini} \; ,
\end{equation}
so, it is a positive quantity and it tells us that 
\begin{equation}
 \frac{a_{mod}^{end}}{a_{mod}^{ini}} = 10^{\Delta x} \; .
\end{equation}
If we fix $x_{mod}^{end}$ to the value $x_{bbn} - 1$, then, as we decrease $x_{mod}^{ini}$ from $x_{mod}^{end} - 1$ 
to $x_{mod}^{end} - 15$, $\Delta x$ changes from $1$ to $15$. This means that the ratio $a_{mod}^{end}/a_{mod}^{ini}$ changes from $10$ to $10^{15}$.
We found that the corresponding value of $H_{\inf}$ changes from 0.442 GeV for $\Delta x = 1$ to 96.79 GeV for $\Delta x = 15$. 
Thus, even though we have introduced a phase of moduli domination in which the Universe expands by a factor of $10^{15}$ i.e. 34.54 e-foldings, the increase in $H_{\inf}$ (determined using TCC) is only a factor of 218.8 (see fig (\ref{Hvsdx})). 
We thus find that as compared to the case in which post-inflationary history is purely radiation dominated, the TCC compatible $H_{\inf}$ is two orders of magnitude higher in the case in which the post inflationary history has a matter dominated phase of sufficiently large duration.
However, even when this enhancement is taken into account, $H_{\inf}$ does not become larger than $m_{\rm mod} \gsim 10^4$ GeV and this is crucial (see the discussion in \textsection \ref{sec:remind}).  
 
\subsubsection{Evolution of $H$ during inflation?}

Before proceeding, we'd like to address possible concerns about the assumption that the Hubble parameter during inflation is taken to be exactly constant. This is correct only approximately and hence one might wonder whether doing a more careful analysis with inclusion of time dependence of $H$ will change the conclusions. 
Firstly, unless the corresponding increase in the value of TCC compatible $H_{\inf}$ is by more than another factor of $10^3$, $H_{\inf}$ shall still not become large compared to $m_{\rm mod}$.
Secondly, the Hubble parameter can not increase with time (unless one is ready to violate the null energy condition), so it must decrease during as inflation proceeds. This means that $y_{mod}(x)$ in fig (\ref{evol-simple}) will no longer be a straight horizontal line but a curve for which $y$ values will increase as we increase $x$. This will imply that $x_{\inf}^0$ will be lower (and hence shift to the left). Thus, a line parallel to the blue line and passing through the point with new values of the coordinates $(x_{\inf}^0, y_{min})$ will be above the red curve: this will violate TCC (for the chosen value of $x_{\inf}^{end}$). In order to satisfy TCC, we would then need to use a higher value of $x_{\inf}^{end}$ resulting in a higher value of $y_{mod}(x_{\inf}^{end})$ and hence a decrease in the inferred value of $H_{\inf}$.

So, the conclusions of this section are: 
(a) the introduction of moduli dominated phase enhances the inferred value of TCC consistent $H_{\inf}$ as compared to a case in which post-inflationary history is purely thermal, 
(b) this enhancement is by a factor of $10^2$ for a duration of matter dominated phase which lasts for 35 e-foldings,
(b) however, this increase is not large enough to displace the moduli from the minima of their potentials.

\subsection{An approximate analytical way of understanding the result} \label{sec:analytical}

The result we obtained seems a little surprising because a casual look at table \ref{table:summary_inf} seems to suggest that having a modulus dominated phase, with slope 3/2 should lead to a TCC compatible value which is in between $10^{10}$ GeV and $10^{-1}$ GeV, while we get $10^2$ GeV. Thus, it is important to try to gain a better understanding of why the enhancement is not large enough, we do this by using rough analytical arguments.

\begin{figure*} 
 \includegraphics[width = .85\textwidth]{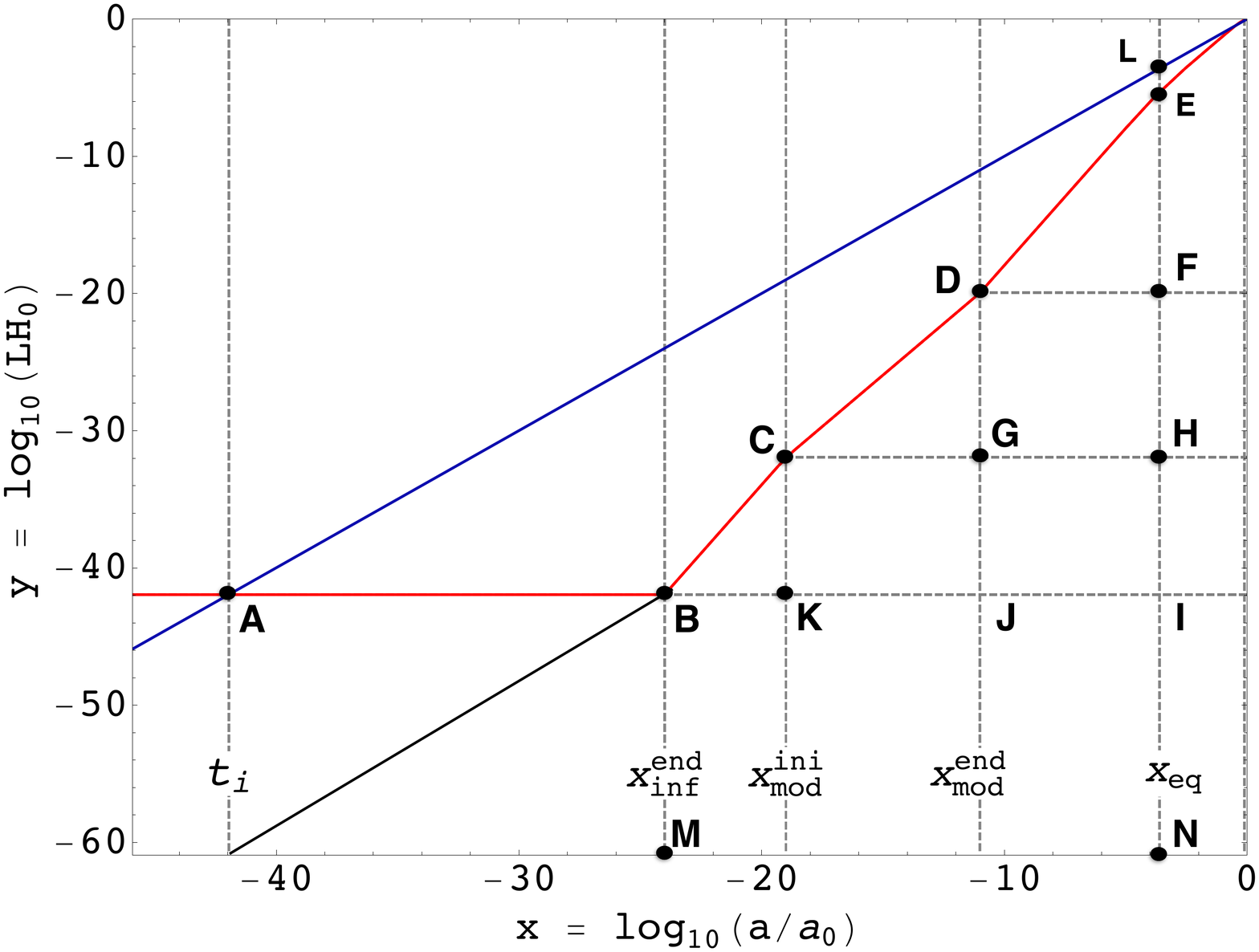}
   \caption{
   An simplified version of fig (\ref{evol-simple}) useful for the analytical arguments of section \textsection \ref{sec:analytical}.
   }
 \label{fig-approx}  
\end{figure*}

We define $N_{\rm inf}$ to be the number of e-folds during inflation. The TCC tells that 
   \begin{equation}
   \frac{H_{\rm inf}}{M_{pl}} \le e^{-N_{\rm inf}}. 
 \end{equation}
 If one considers the post inflationary Universe to be only radiation dominated epoch till BBN,  one can get~(see cite{Shinji})
 \beq
 \label{eq:Ninf}
 N_{\rm inf} \sim \frac{1}{3}{\rm ln}\frac{M_{pl}}{H_0}.
 \eeq
  Using TCC constraint, it gives $H_{\rm inf} < 0.1~{\rm GeV}$. 
 
 Now we try to estimate the change in $N_{\rm inf}$ due to an intermediate matter dominated regime present in the post-inflationary history of the universe by using simple trigonometry in fig (\ref{fig-approx}). We  define
 \begin{enumerate}
 \item  $N_{\rm rh} = x^{\rm ini}_{\rm mod} - x^{\rm end}_{\rm inf}$,  number of e-folds between the end of inflation and the beginning of radiation dominated era.
 \item 
  $N_{\rm mod, rh} = x^{\rm end}_{\rm mod} - x^{\rm ini}_{\rm mod}$, number of e-folds between the beginning and the end of the modulus dominated era. 
  \item $N_{\rm eq} = x_{\rm eq} -  x^{\rm end}_{\rm mod}$, number of e-folds between the end of the modulus dominated era and present time.
 \end{enumerate}

By using the last column of table I, we can see in fig (\ref{fig-approx}):
\begin{enumerate}
\item  In ${\Delta{\rm BCK}}$ (corresponding to radiation dominated era), slope of BC = 2, implying {\rm HI} = ${\rm CK} = 2{\rm BK} = 2 N_{\rm rh}$.  
\item  In ${\Delta{\rm CDG}}$ (corresponding to matter dominated era), slope of CD = $\frac{3}{2}$, implying {\rm FH} = ${\rm DG} = \frac{3}{2} {\rm CG} =  \frac{3}{2}N_{\rm mod, rh}$.
\item In ${\Delta{\rm DEF}}$ (corresponding to radiation dominated era), slope of DE = 2, implying ${\rm EF} = 2{\rm DF} = 2 N_{\rm eq}$.
\item Since number of e-folds between the time of matter-radiation equality and the present time is very small, we assume  point L to be sitting on top of E in fig (\ref{fig-approx}). Thus, for ${\Delta{\rm AEI}}$, we have slope of AE = 1, implying ${\rm EI} = {\rm AI}$. 
\end{enumerate}
In the last case,  ${\rm EI} = {\rm AI}$ corresponds to ${\rm EF} + {\rm FH} + {\rm HI}  = {\rm AB} + {\rm BK} +  {\rm KJ} +  {\rm JI}$. By simplifying this, we get
 \begin{equation}
 \label{eq.Neq}
N_{\rm eq} = N_{\rm inf} - N_{\rm rh} - \frac{N_{\rm mod, rh}}{2},
 \end{equation}
Next, we have 
    \begin{equation}
   \label{eq.totN1}
 {\rm EN} = {\rm EF} + {\rm FH} + {\rm HI} + {\rm IN}  = {\rm ln}\frac{M_{pl}}{H_0}.
 \end{equation}
By using eqs.~(\ref{eq.Neq}) and (\ref{eq.totN1}), we get
   \begin{equation}
   \label{eq.totN}
  3 N_{\rm inf} + \frac{N_{\rm mod, rh}}{2} = {\rm ln}\frac{M_{pl}}{H_0}.
 \end{equation}
Finally, by adding eqs.~(\ref{eq.Neq}) and (\ref{eq.totN}), we get
   \begin{equation}
   \label{eq.finN}
  N_{\rm inf} =   \frac{1}{4}\left[{\rm ln}\frac{M_{pl}}{H_0} +  N_{\rm rh} + N_{\rm eq} \right].
 \end{equation}
 We can calculate $N_{\rm eq}$ by considering $N_{\rm eq} = {\rm ln}\left(\frac{a_0}{a^{\rm end}_{\rm mod}}\right) \sim {\rm ln}\left(\frac{T^{\rm mod}_{\rm rh}}{T_0}\right)$, where $T^{\rm mod}_{\rm rh}$ is the reheating temperature of the universe after the decay of the moduli and $T_0$ is the present temperature.
Similarly, we can  take $N_{\rm rh} \sim {\rm ln}\left(\frac{T^{\rm inf}_{\rm rh}}{T^{\rm mod}_{\rm rh}}\right)$, where $T^{\rm inf}_{\rm rh}$ is the reheating temperature of the universe after the end of inflation. 
 By comparing eqs.~(\ref{eq:Ninf}) and (\ref{eq.finN}), we can see that there is negligible change in the value of $N_{\rm inf}$ due to presence of intermediate moduli dominated era in the post-inflationary history of the universe. 
 Of course, the more accurate numerical analysis that we did earlier suggests a little change, but this change is small. This explains the numerical results of the previous sections.

\section{Discussion} \label{sec:discussion}

Careful studies of explicit string compactifications as well as some very general arguments lead to the possibility that low energy effective field theories arising from consistent theories of quantum gravity (which belong to string landscape) could be distinguishable from arbitrary quantum field theories (which belong to the swamplnad) \cite{Brennan:2017rbf}, \cite{Danielsson:2018ztv}, \cite{Palti:2019pca}, \cite{Roupec:2018mbn}.
Trans-Planckian Censorship Conjecture is one such conjectured property of solutions which arise in consistent theories of quantum gravity.

But typically, string compactifications also have geometric moduli, axions and other open string moduli (e.g. brane positions etc). Studies of the evolution of such moduli fields during and after inflation have suggested that one could have a matter dominated era in the post-inflationary history of the Universe before BBN. Given this, it is extremely important to try to see whether this conclusion gets affected by the Trans-Planckian Censorship Conjecture (TCC). TCC can lead to determination of the energy scale of inflation but the scale implied by TCC depends on the details of the dynamics of post-inflationary universe. We thus tried to find the energy scale of inflation for a situation in which the post-inflationary universe is matter dominated. We found that, as compared to $H_{\rm inf}$ obtained assuming the standard post-inflationary thermal history, the value of $H_{\rm inf}$ obtained in the case of matter domination phase between inflation and BBN, can be two orders of magnitude higher (depending on the number of e-foldings of evolution of the universe during moduli domination). In particular, we found that if moduli domination ended at a time when the scale factor of the universe was 10 times of its value during BBN, and if it lasted for about 35 e-foldings before that, the value of $H_{\rm inf}$ inferred from TCC would be enhanced by $ \approx 200$.

On the other hand, it has been well known that for moduli with gravitational couplings, the requirement that reheating temperature must be higher than  the temperature of BBN implies that the the moduli masses must be more than about 10 TeV. Unless the moduli are lighter than $H_{\rm inf}$, they will not be displaced from their minimum during inflation and there will be no matter dominated phase in the post-inflationary history of the universe before BBN. 

We thus find that, the increase in the value of $H_{\rm inf}$ due to the introduction of post-inflationary matter dominated phase is not large enough to displace the moduli from the bottom of their potentials and so, there can not be a moduli dominated phase. It thus seems that if TCC is correct, and hence the energy scale of inflation is determined by the requirement of saturation of TCC, then, there can be no moduli dominated phase in the early history of the Universe.

\end{document}